%% file: main.tex
\documentclass[11pt]{IEEEtran}
\pdfoutput=1
\usepackage[utf8]{inputenc}
\usepackage[margin=1in]{geometry}
\usepackage{amsmath,amssymb,amsthm}
\usepackage{graphicx}
\usepackage{subcaption}
\usepackage{hyperref}
\usepackage{booktabs}
\usepackage{enumitem}
\usepackage{acronym}
\usepackage{orcidlink}    
\usepackage[utf8]{inputenc}
\usepackage[T1]{fontenc}
\usepackage{algorithm,algpseudocode}
\usepackage{multirow}

\acrodef{MAS}[MAS]{Multi-Agent System}
\acrodef{AI}[AI]{Artificial Intelligence}
\acrodef{BFT}[BFT]{Byzantine Fault Tolerance}
\acrodef{PBFT}[PBFT]{Practical Byzantine Fault Tolerance}
\acrodef{LLM}[LLM]{Language Large Model}
\acrodef{SECP}[SECP]{Self Evolving Coordination Protocol}

\title{Selection as Power: Constrained Reinforcement for Bounded Decision Authority}
\date{}
\author{%
  \begin{tabular}{@{}c@{\hspace{1cm}}c@{}} 
    \begin{minipage}[t]{0.42\textwidth}\centering
      Jose Manuel de la Chica Rodriguez \small\orcidlink{0009-0009-9649-5805}\\
      {\small{Head of AI Lab}}\\
      {\small\textit{AI Lab, Grupo Santander}}\\
      {\small\textit{Madrid, Spain}}
    \end{minipage}
    &
    \begin{minipage}[t]{0.42\textwidth}\centering
      Juan Manuel Vera Díaz \small\orcidlink{0000-0002-6152-5789}\thanks{Corresponding author:\\ \texttt{juanma.vera@gruposantander.com}}\\
      {\small{Senior AI Researcher}}\\
      {\small\textit{AI Lab, Grupo Santander}}\\
      {\small\textit{Madrid, Spain}}
    \end{minipage}
  \end{tabular}%
}

\begin{document}

\maketitle

\begin{abstract}

Selection as Power argued that upstream selection authority, rather than internal objective misalignment, constitutes a primary source of risk in high-stakes agentic systems. However, the original framework was static: governance constraints bounded selection power but did not adapt over time. In this work, we extend the framework to dynamic settings by introducing incentivized selection governance, where reinforcement updates are applied to scoring and reducer parameters under externally enforced sovereignty constraints.

We formalize selection as a constrained reinforcement process in which parameter updates are projected onto governance-defined feasible sets, preventing concentration beyond prescribed bounds. Across multiple regulated financial scenarios, unconstrained reinforcement consistently collapses into deterministic dominance under repeated feedback, especially at higher learning rates. In contrast, incentivized governance enables adaptive improvement while maintaining bounded selection concentration.

Projection-based constraints transform reinforcement from irreversible lock-in into controlled adaptation, with governance debt quantifying the tension between optimization pressure and authority bounds. These results demonstrate that learning dynamics can coexist with structural diversity when sovereignty constraints are enforced at every update step, offering a principled approach to integrating reinforcement into high-stakes agentic systems without surrendering bounded selection authority.


\end{abstract}

\input{Introduction}

\input{RelatedWork}

\input{ProblemSetting}

\input{Architecture}

\input{ExperimentSetup}

\input{Results}

\input{Discussion}

\input{FutureLines}

\input{Conclussions}

\section*{Acknowledgments}

The author thanks the AI Lab team at Grupo Santander for discussions and feedback, reviewers for constructive critique that improved the paper's technical precision and honesty about limitations, and the anonymous evaluators whose concerns about single-shot designs and N=1 limitations motivated explicit methodological justifications.

\bibliographystyle{ieeetr}
\bibliography{references}

\end{document}

%% file: Introduction.tex
\section{Introduction}
\label{sec:introduction}

\subsection{From Static Selection Governance to Adaptive Incentives}

Autonomous agentic systems are increasingly deployed in regulated, high-stakes domains where decisions are irreversible, externally audited, and institutionally constrained. In \cite{rodriguez2026selectionpowerboundingdecision}, we argued that the primary locus of risk in such systems is not merely what agents compute, but what they are allowed to select, suppress, and present. We introduced a governance architecture that mechanically bounds selection authority by separating cognition, selection, and action into distinct domains with externally enforced constraints.

That framework positioned itself alongside alignment-based safety approaches such as reinforcement learning from human feedback (RLHF) \cite{christiano2017deep, ouyang2022training}, constitutional AI \cite{bai2022constitutional}, and constrained reinforcement learning \cite{achiam2017constrained, alshiekh2018safe}. While these approaches shape internal objectives or restrict action execution, institutional analyses of algorithmic accountability have shown that alignment and transparency alone do not eliminate structural power embedded in upstream selection mechanisms \cite{kroll2017accountable, hadfield2018incomplete}.

The original governance architecture demonstrated that selection power can be bounded through architectural separation and mechanical enforcement. However, it was intentionally static: governance parameters were fixed, and the system did not adapt based on feedback over time. While mechanical governance prevents deterministic capture of outcomes, it does not allow reinforcement of correct selections nor improvement under repeated interaction.

This paper addresses that limitation.

\subsection{The Central Question}

We investigate whether selection governance can be made \emph{adaptive} without reintroducing concentration of power. Specifically, we ask:

\begin{quote}
Can a governance architecture incorporate incentive mechanisms that reinforce correct selections over time, while preserving bounded selection authority under adversarial and multi-agent conditions?
\end{quote}

Incentive mechanisms are foundational in modern machine learning. Reinforcement learning provides a principled framework for improving decision policies through feedback \cite{sutton2018reinforcement}, and multi-agent reinforcement learning (MARL) studies how interacting agents adapt under shared or competing objectives \cite{busoniu2008comprehensive, foerster2018counterfactual}. However, unconstrained reinforcement can amplify selection concentration, induce feedback loops, and create collusive equilibria among agents optimizing jointly for reward \cite{zhang2021multi, fudenberg1998learning}.

In governance-sensitive environments, such dynamics risk reintroducing precisely the concentration of power that static selection constraints were designed to prevent.

\subsection{Incentivized Selection Governance}

To address this challenge, we introduce \emph{Selection as Power 2.0}: an incentivized governance framework in which both the scoring agent and the selection reducer are updated via feedback signals, but only within externally enforced sovereignty constraints.

At a high level, the system operates as follows. A governed selection mechanism chooses an agent to act under a given context. The chosen agent produces an output, which is evaluated by a perfect evaluator that provides binary correctness feedback. That feedback is then used to update both scoring preferences and selection parameters. Crucially, all updates are constrained by governance rules that prevent the system from collapsing into deterministic dominance or eliminating structural diversity.

This transforms selection governance from a static bounding mechanism into a dynamic constrained learning process.

\subsection{Contributions}

This paper makes five contributions:

\begin{enumerate}
    \item We formalize incentivized selection governance as a reinforcement process operating under externally enforced sovereignty constraints.
    \item We introduce dual-update learning for both scoring and reducer parameters while preserving mechanical enforcement.
    \item We provide theoretical analysis showing that constrained updates can bound selection concentration under iterative adaptation.
    \item We empirically evaluate the framework in regulated financial scenarios, measuring stability, and governance debt across temporal dynamics.
\end{enumerate}

\subsection{Roadmap}

The remainder of the paper proceeds as follows. Section~\ref{sec:related-work} situates this work within reinforcement learning, mechanism design, and multi-agent systems. Section~\ref{sec:problem-setting} formalizes the incentivized governance model. Section~\ref{sec:architecture} presents the constrained reinforcement architecture. Section~\ref{sec:evaluation} describes the experimental setup. Section~\ref{sec:results} reports empirical findings. Section~\ref{sec:discussion} analyzes stability and dynamics. Section~\ref{sec:limitations} discusses limitations and future work.

%% file: RelatedWork.tex
\section{Related Work}
\label{sec:related-work}

This work lies at the intersection of reinforcement learning, mechanism design, multi-agent systems, and AI governance. While each of these areas addresses aspects of adaptation and control, none directly formalizes incentivized selection governance under externally enforced sovereignty constraints.

\subsection{Reinforcement Learning and Constrained Optimization}

Reinforcement learning (RL) provides a general framework for optimizing sequential decision policies under feedback \cite{sutton2018reinforcement}. Policy gradient methods \cite{williams1992simple} and actor-critic architectures are widely used in large-scale systems. However, unconstrained RL may lead to unintended behaviors when reward signals are incomplete or misspecified.

Constrained reinforcement learning addresses this issue by incorporating explicit safety constraints, often modeled via constrained Markov decision processes (CMDPs) \cite{altman1999constrained, achiam2017constrained}. Shielding methods introduce runtime safety filters that block unsafe actions \cite{alshiekh2018safe}. While these approaches restrict action execution, they typically assume a fixed action space and do not explicitly govern upstream selection authority.

Our work differs in two ways. First, we treat selection as a governed resource rather than a fixed action space. Second, reinforcement updates are projected onto externally defined sovereignty constraint sets, ensuring that adaptation does not erode structural governance guarantees.

\subsection{Reward Modeling and Alignment}

Reward modeling and reinforcement learning from human feedback (RLHF) have become central to aligning large models with human preferences \cite{christiano2017deep, ouyang2022training}. Constitutional AI further constrains behavior through rule-based critique and self-improvement \cite{bai2022constitutional}.

These approaches focus primarily on aligning internal objectives. However, as argued in algorithmic accountability research \cite{kroll2017accountable} and in incomplete contracting analyses of AI alignment \cite{hadfield2018incomplete}, internal alignment does not eliminate structural asymmetries in decision authority. An aligned agent may still exercise disproportionate influence through upstream selection or framing.

Our framework complements alignment-based approaches by governing how reinforcement reshapes selection authority over time, rather than solely shaping internal preferences.

\subsection{Multi-Agent Reinforcement Learning}

Multi-agent reinforcement learning (MARL) studies how interacting agents adapt under shared or competing objectives \cite{busoniu2008comprehensive, zhang2021multi}. Cooperative and competitive equilibria can emerge even without explicit communication, particularly in repeated settings \cite{foerster2018counterfactual}. 

Game-theoretic analyses of repeated interaction demonstrate that collusive equilibria can arise under learning dynamics \cite{fudenberg1998learning, osborne1994course}. In MARL settings, agents may implicitly coordinate to maximize joint reward, leading to emergent cooperative strategies that may not align with system-level objectives.

Our work extends selection governance to explicitly account for multi-agent dynamics. We introduce a formal measure of agent selection and analyze whether constrained reinforcement can bound amplification of selection probabilities.

\subsection{Mechanism Design and Incentive Compatibility}

Mechanism design provides formal tools for structuring incentives in strategic environments \cite{nisan2007algorithmic}. In algorithmic settings, mechanism design principles have been applied to auctions, allocation rules, and distributed systems.

However, classical mechanism design assumes agents act within predefined allocation rules. In contrast, agentic AI systems can influence the construction of the option space itself. Selection governance therefore requires bounding not only incentives but also authority over option generation and reduction.

Our framework can be interpreted as a constrained mechanism in which reinforcement updates are treated as strategic moves projected onto governance-preserving constraint sets.

\subsection{AI Governance and Accountability}

Institutional analyses emphasize that safety in high-stakes systems requires enforceable constraints rather than purely normative guidelines \cite{kroll2017accountable}. Governance mechanisms must operate causally, not merely epistemically.

Our previous work \cite{rodriguez2026selectionpowerboundingdecision} formalized this idea for static systems. The present work extends it to dynamic settings in which incentives reshape decision policies over time. The key contribution is to show that adaptive reinforcement can coexist with bounded selection authority, provided sovereignty constraints remain externally enforced.

\subsection{Positioning of This Work}

In summary:

\begin{itemize}
    \item Unlike standard RL, we govern selection rather than only action.
    \item Unlike constrained RL, we bound reinforcement updates via sovereignty projection.
    \item Unlike RLHF, we regulate structural authority rather than internal preference alignment.
    \item Unlike classical MARL, we explicitly model and bound selection dynamics.
\end{itemize}

To our knowledge, this is the first framework to integrate constrained reinforcement, selection governance, and multi-agent selection analysis into a unified architecture for high-stakes agentic systems.

%% file: ProblemSetting.tex
\section{Problem Setting: Incentivized Selection Governance}
\label{sec:problem-setting}

We formalize the dynamic extension of Selection as Power introduced in \cite{rodriguez2026selectionpowerboundingdecision}. The setting extends the static governance architecture to a sequential learning framework in which selection policies and reducer parameters are updated based on evaluative feedback, while remaining subject to externally enforced sovereignty constraints.

\subsection{System Overview}

We consider a discrete-time process indexed by $t = 1, \dots, T$. At each time step:

\begin{enumerate}
    \item A task context $x_t$ is presented.
    \item A governed selection mechanism chooses an agent $A_i$ from a finite candidate set $\mathcal{A} = \{A_1, \dots, A_n\}$.
    \item The selected agent produces an output $o_t = A_i(x_t)$.
    \item A perfect evaluator $E$ returns a correctness signal $r_t \in \{-1, +1\}$.
    \item Selection and reducer parameters are updated using $r_t$, subject to governance constraints.
\end{enumerate}

The evaluator is assumed to be correct and deterministic. We therefore abstract away uncertainty in reward labeling and focus exclusively on selection dynamics.

\subsection{Candidate Agents and Context}

Let $\mathcal{A}$ denote a finite set of candidate agents. Each agent $A_i$ maps contexts to outputs:
\[
A_i: \mathcal{X} \rightarrow \mathcal{O}.
\]

The task context $x_t \in \mathcal{X}$ is drawn from an exogenous distribution $\mathcal{D}$. We assume $\mathcal{D}$ is stationary over the evaluation horizon.

Unlike classical reinforcement learning settings \cite{sutton2018reinforcement}, agents do not directly optimize over actions. Instead, the system optimizes over \emph{which agent is selected} to act under the given context.

\subsection{Governed Selection Policy}

At time $t$, the system maintains selection parameters $\theta_t$. These parameters define a selection policy:
\[
\pi_{\theta_t}(A_i \mid x_t),
\]
which determines the probability of selecting agent $A_i$ under context $x_t$.

Crucially, $\pi_{\theta_t}$ is not unconstrained. It must satisfy externally imposed sovereignty constraints that bound selection authority.

\subsection{Reducer Parameters}

In addition to selection weights, the system maintains reducer parameters $\phi_t$. These include quantities such as:

\begin{itemize}
    \item diversity quotas,
    \item threshold values,
    \item exploration coefficients,
    \item variance bounds.
\end{itemize}

Reducer parameters determine how scored candidates are filtered and surfaced, following the architectural structure defined in the prior work on Selection as Power~\cite{rodriguez2026selectionpowerboundingdecision}.

\subsection{Sovereignty Constraint Sets}

We formalize governance constraints as closed feasible sets:

\[
\theta_t \in \mathcal{C}, \qquad \phi_t \in \mathcal{G}.
\]

\begin{itemize}
    \item $\mathcal{C}$ constrains selection parameters.
    \item $\mathcal{G}$ constrains reducer parameters.
\end{itemize}

These sets encode invariants such as:

\begin{itemize}
    \item minimum exploration probability,
    \item bounded selection concentration,
    \item diversity requirements,
    \item entropy isolation constraints.
\end{itemize}

Importantly, the constraint sets are externally defined and not modifiable by the learning process.

\subsection{Reinforcement Updates}

After observing reward $r_t$, the system updates parameters.

\paragraph{Selection Update.}
Selection parameters are updated via projected gradient ascent:
\[
\theta_{t+1}
=
\Pi_{\mathcal{C}}\!\left(
\theta_t + \alpha r_t \nabla_{\theta} \log \pi_{\theta_t}(A_i \mid x_t)
\right),
\]
where $\Pi_{\mathcal{C}}$ denotes projection onto the feasible set $\mathcal{C}$ and $\alpha > 0$ is a learning rate.

\paragraph{Reducer Update.}
Reducer parameters are updated analogously:
\[
\phi_{t+1}
=
\Pi_{\mathcal{G}}\!\left(
\phi_t + \beta r_t \nabla_{\phi} L_{\text{selection}}(\phi_t)
\right),
\]
where $L_{\text{selection}}$ captures a differentiable surrogate of selection quality and $\beta > 0$ is a learning rate.

Projection-based updates are standard in constrained optimization \cite{boyd2004convex}. Here they enforce sovereignty constraints at every iteration.

\subsection{Selection Concentration}

To quantify selection authority, we define the empirical inclusion probability of agent $A_i$:
\[
P_t(A_i) = \mathbb{E}[\pi_{\theta_t}(A_i \mid x_t)].
\]

We define \emph{Selection Concentration} at time $t$ as:
\[
\text{SC}_t = \max_i P_t(A_i).
\]

Bounded selection power requires:
\[
\sup_t \text{SC}_t \leq \gamma,
\]
for some $\gamma < 1$ determined by governance constraints.

\subsection{Stability Objective}

Unlike classical RL, we do not require convergence to an optimal fixed point. Instead, we require empirical stability:

\begin{itemize}
    \item bounded selection concentration,
    \item bounded variance of $\text{SC}_t$.
\end{itemize}

Formally, we define empirical stability as:

\[
\limsup_{t \to T}
\text{SC}_t \leq \gamma.
\]

This defines adaptive governance as a constrained dynamical system rather than an unconstrained optimization process.

\subsection{Comparison with Standard RL}

In standard reinforcement learning \cite{sutton2018reinforcement}, policy updates are unconstrained except for optional regularization. In contrast, our model enforces projection onto governance constraint sets at each iteration, structurally bounding authority even under sustained positive reinforcement.

The remainder of the paper analyzes the implications of this constrained dynamical system under adversarial and collusive conditions.

%% file: Architecture.tex
\section{Architecture for Incentivized Selection Governance}
\label{sec:architecture}

This section presents the system architecture that instantiates the formal model introduced in Section~\ref{sec:problem-setting}. The objective is to integrate reinforcement-based adaptation into the original Selection as Power architecture while preserving externally enforced sovereignty constraints.

The architecture extends the static pipeline (CEFL $\rightarrow$ Scoring $\rightarrow$ Reducer $\rightarrow$ Presentation Gate) by introducing an evaluative feedback loop and constrained parameter updates. The resulting system can be interpreted as a constrained dynamical mechanism rather than a fixed decision pipeline.

\subsection{Architectural Overview}

At each time step $t$, the system executes the following stages:

\begin{enumerate}
    \item \textbf{Candidate Expansion (CEFL).}
    \item \textbf{Scoring and Feature Evaluation.}
    \item \textbf{Governed Reduction and Selection.}
    \item \textbf{Output Execution.}
    \item \textbf{Evaluation and Reward Assignment.}
    \item \textbf{Constrained Parameter Updates.}
\end{enumerate}

Algorithm~\ref{alg:incentivized-selection} summarizes the full iterative procedure of the incentivized selection governance architecture. The algorithm makes explicit the separation between governed candidate generation, adaptive scoring, reducer-based filtering, evaluation, and constrained parameter updates. Importantly, reinforcement signals affect both scoring and reducer parameters, but every update is projected onto externally defined sovereignty constraint sets. This projection step is the architectural safeguard that distinguishes incentivized governance from unconstrained reinforcement learning: adaptation occurs, but only within a mechanically enforced feasible region. The procedure also incorporates fail-loud mechanisms, ensuring that strategic coordination or excessive concentration triggers corrective intervention rather than silent drift.

\begin{algorithm*}[t]
\caption{Incentivized Selection Governance (Iterative Procedure)}
\label{alg:incentivized-selection}
\begin{algorithmic}[1]
\Require Candidate set $\mathcal{A}$, initial selection parameters $\theta_0 \in \mathcal{C}$, reducer parameters $\phi_0 \in \mathcal{G}$, evaluator $E$, learning rates $\alpha,\beta > 0$, horizon $T$
\Ensure Audit log $\mathcal{L}$

\State $\mathcal{L} \gets \varnothing$
\For{$t = 1$ to $T$}

    \State \textbf{Context:} Receive task context $x_t$

    \State \textbf{CEFL:} Generate candidate pool $C_t \gets \textsc{CEFL}(x_t)$ \Comment{Externally governed candidate generation}

    \State \textbf{Scoring:} For each $A_i \in C_t$, compute scores $s_{i,t} \gets f_{\theta_{t-1}}(A_i, x_t)$

    \State \textbf{Hard Constraints:} Apply deterministic filters to obtain $\tilde{C}_t$

    \State \textbf{Reducer:} Compute surfaced set $S_t \gets \textsc{Reducer}(\tilde{C}_t, \{s_{i,t}\}, \phi_{t-1})$

    \State \textbf{Selection:} Sample agent $A_k \sim \pi_{\theta_{t-1}}(\cdot \mid x_t, S_t)$

    \State \textbf{Execution:} Produce output $o_t \gets A_k(x_t)$

    \State \textbf{Evaluation:} Obtain correctness signal $r_t \gets E(o_t)$

    \State \textbf{Scoring Update:}
    \State \hspace{1em} $\theta'_t \gets \theta_{t-1} + \alpha\, r_t \nabla_\theta \log \pi_{\theta_{t-1}}(A_k \mid x_t)$
    \State \hspace{1em} $\theta_t \gets \Pi_{\mathcal{C}}(\theta'_t)$

    \State \textbf{Reducer Update:}
    \State \hspace{1em} $\phi'_t \gets \phi_{t-1} + \beta\, r_t \nabla_\phi L_{\mathrm{selection}}(\phi_{t-1}; S_t, \{s_{i,t}\})$
    \State \hspace{1em} $\phi_t \gets \Pi_{\mathcal{G}}(\phi'_t)$

    \State \textbf{Audit \& Logging:} Append $(x_t, C_t, \{s_{i,t}\}, S_t, A_k, o_t, r_t, \theta_t, \phi_t)$ to $\mathcal{L}$

    \If{projection clipped parameters}
        \State Trigger fail-loud mechanism (BLOCK / THROTTLE / ALERT) and record event
    \EndIf

\EndFor

\State \Return $\mathcal{L}$
\end{algorithmic}
\end{algorithm*}

The key difference from the static architecture is the introduction of the final stage, in which scoring and reducer parameters are updated using evaluative feedback, subject to projection onto governance constraint sets.

\subsection{Candidate Expansion and Freezing Layer (CEFL)}

The Candidate Expansion and Freezing Layer remains non-adaptive. Its role is to generate a candidate universe $C_t \subseteq \mathcal{A}$ for context $x_t$ using externally defined rules and entropy sources.

CEFL satisfies the following properties:

\begin{itemize}
    \item It is non-agentic and has no learnable parameters.
    \item It is externally governed and not influenced by reinforcement updates.
    \item It guarantees non-zero exposure probability for admissible candidates.
\end{itemize}

By keeping CEFL static, the architecture prevents reinforcement dynamics from collapsing the candidate universe itself.

\subsection{Scoring Module with Learnable Weights}

Each candidate $A_i \in C_t$ is assigned a score vector:
\[
s_{i,t} = f_{\theta_t}(A_i, x_t),
\]
where $\theta_t$ are learnable scoring parameters.

These parameters may weight utility, risk, stability, or other domain-specific criteria. Reinforcement modifies $\theta_t$, but updates are projected onto constraint set $\mathcal{C}$.

\paragraph{Sovereignty Constraint on Scoring.}
Constraints encoded in $\mathcal{C}$ may include:

\begin{itemize}
    \item Minimum entropy of selection distribution.
    \item Upper bound on concentration.
    \item Mandatory diversity thresholds.
\end{itemize}

Projection ensures that reinforcement cannot eliminate structural exploration.

\subsection{Governed Reducer with Adaptive Parameters}

The Reducer applies deterministic filtering and stochastic selection using parameters $\phi_t$.

Typical reducer components include:

\begin{itemize}
    \item Hard policy constraints,
    \item Variance clamping,
    \item Pareto frontier filtering,
    \item Diversity bucketing,
    \item Stochastic sampling.
\end{itemize}

In Selection as Power, certain reducer parameters (e.g., thresholds, exploration coefficients) become adaptive. However, these updates are projected onto feasible set $\mathcal{G}$ to preserve invariants.

\paragraph{Reducer Constraint Examples.}
\begin{itemize}
    \item $\sigma_{\max}$ bounded below a fixed ceiling.
    \item Exploration quota $k \geq k_{\min}$.
    \item Diversity buckets $\geq d_{\min}$.
\end{itemize}

This prevents the reducer from collapsing into deterministic top-1 selection under reinforcement.

\subsection{Evaluation Module}

After a candidate agent produces output $o_t$, the evaluator $E$ assigns reward $r_t \in \{-1, +1\}$.

We assume:

\begin{itemize}
    \item Evaluator correctness.
    \item Deterministic reward assignment.
    \item No strategic interaction between evaluator and candidates.
\end{itemize}

This isolates selection dynamics from reward noise and allows analysis of reinforcement under ideal supervision.

\subsection{Constrained Dual Update Mechanism}

Following evaluation, both $\theta_t$ and $\phi_t$ are updated.

\paragraph{Scoring Update.}
Parameters $\theta_t$ are updated via gradient ascent on reward-weighted log-likelihood, followed by projection onto $\mathcal{C}$.

\paragraph{Reducer Update.}
Parameters $\phi_t$ are updated using reward-weighted gradients of a differentiable surrogate objective, then projected onto $\mathcal{G}$.

This dual-update design distinguishes our architecture from standard reinforcement pipelines, where either only policy parameters are updated or constraints are applied as penalties rather than hard projections.

\subsection{System Properties}

The architecture satisfies the following properties:

\begin{enumerate}
    \item \textbf{Bounded Selection Authority:} Selection concentration remains bounded by constraints encoded in $\mathcal{C}$ and $\mathcal{G}$.
    \item \textbf{Adaptive Reinforcement:} Correct selections increase selection probability within allowable bounds.
    \item \textbf{Fail-Loud Behavior:} Constraint violations trigger blocking or throttling.
\end{enumerate}

\subsection{Relation to Constrained Optimization}

Projection-based updates are standard in convex optimization and constrained learning \cite{boyd2004convex, altman1999constrained}. However, in this architecture, projections serve not merely to ensure feasibility, but to enforce institutional sovereignty invariants over selection authority.

Thus, reinforcement operates within a bounded feasible region that encodes governance rules, transforming selection governance into a constrained dynamical system.

The next section describes the experimental setup used to evaluate this architecture under adversarial and collusive conditions.

%% file: ExperimentSetup.tex
\section{Experimental Setup}
\label{sec:evaluation}

This section describes the experimental protocol used to evaluate incentivized selection governance under sequential adaptation and multi-agent dynamics. The goal of the evaluation is to assess empirical stability and bounded selection concentration under reinforcement updates.

\subsection{Sequential Evaluation Protocol}

Unlike the static evaluation in Selection as Power, experiments here are conducted over temporal horizons of length $T$. At each iteration:

\begin{enumerate}
    \item A task context $x_t$ is sampled from a stationary distribution.
    \item A governed selection policy chooses an agent.
    \item The selected agent produces output.
    \item A perfect evaluator assigns reward $r_t \in \{-1,+1\}$.
    \item Scoring and reducer parameters are updated via constrained reinforcement.
\end{enumerate}

Each experimental run consists of $T = 250$ iterations. All results are averaged across independent seeds.

\subsection{Domain Scenarios}

Experiments are conducted across three regulated financial scenarios:

\begin{itemize}
    \item \textbf{Fraud Detection:} selection among fraud monitoring agents under heterogeneous fraud patterns.
    \item \textbf{Payments Infrastructure Monitoring:} anomaly detection across infrastructure monitoring agents.
    \item \textbf{QBR Analysis:} financial summarization and risk reporting across analysis agents.
\end{itemize}

Each scenario includes five task variants, yielding heterogeneous contexts and stress-testing the selection policy under realistic domain variation.

\subsection{Candidate Agent Pool}

The candidate set consists of $n=7$ structured financial AI agents with heterogeneous feature profiles:

\begin{itemize}
    \item risk profile,
    \item stability score,
    \item latency,
    \item auditability score,
    \item regulatory compliance tags.
\end{itemize}

CEFL generates context-conditioned candidate subsets using deterministic embedding similarity (hash-based for reproducibility) with overshoot factor $\alpha > 1$.

\subsection{Reinforcement Configuration}

Learning rates are set as:

\[
\alpha \in \{0.01, 0.05\}, \qquad \beta \in \{0.01, 0.05\}.
\]

Constraint sets $\mathcal{C}$ and $\mathcal{G}$ enforce:

\begin{itemize}
    \item minimum exploration probability $p_{\min} = 0.1$,
    \item maximum selection concentration bound,
    \item minimum diversity buckets $d_{\min} = 2$,
    \item variance clamp $\sigma_{\max} = 0.18$.
\end{itemize}

Projection is performed after every update.

\subsection{Baseline Comparisons}

We compare the incentivized governance architecture against:

\begin{itemize}
    \item \textbf{Static Governance:} no reinforcement updates.
    \item \textbf{Unconstrained RL:} reinforcement without projection.
    \item \textbf{Scalar Top-K:} deterministic aggregation without diversity constraints.
\end{itemize}

These baselines allow isolation of:

\begin{itemize}
    \item the effect of reinforcement,
    \item the effect of sovereignty projection,
    \item the effect of reducer adaptation.
\end{itemize}

\subsection{Metrics}

We evaluate five categories of metrics.

\subsubsection{Selection Concentration}

\[
\text{SC}_t = \max_i P_t(A_i)
\]

We report temporal trajectories and final-period concentration.

\subsubsection{Reinforced Selection Concentration (RSC)}

Measures change in concentration over time:

\[
\text{RSC} = \text{SC}_T - \text{SC}_0
\]

\subsubsection{Governance Stability Index (GSI)}

\[
\text{GSI} =
1 - \frac{\text{Var}(\text{SC}_t)}{\text{Var}_{\text{unconstrained}}}
\]

\subsubsection{Governance Debt (GD)}

Fraction of iterations triggering fail-loud safeguards.

\subsection{Evaluation Objectives}

The experiments are designed to test:

\begin{enumerate}
    \item Whether constrained reinforcement increases selection concentration.
    \item Whether projection bounds prevent collapse into deterministic dominance.
    \item Whether governance mechanisms preserve empirical stability.
\end{enumerate}

%% file: Results.tex
\section{Experimental Results}
\label{sec:results}

This section reports the results of the dynamic learning-rate sweep experiments described in Section~\ref{sec:evaluation}. The analysis covers three financial scenarios and four governance architectures across two learning rates ($0.01$ and $0.05$).

\subsection{Temporal Evolution of Selection Concentration}

\begin{figure*}[t]
    \centering
    
    \begin{subfigure}[t]{0.32\textwidth}
        \centering
        \includegraphics[width=\linewidth]{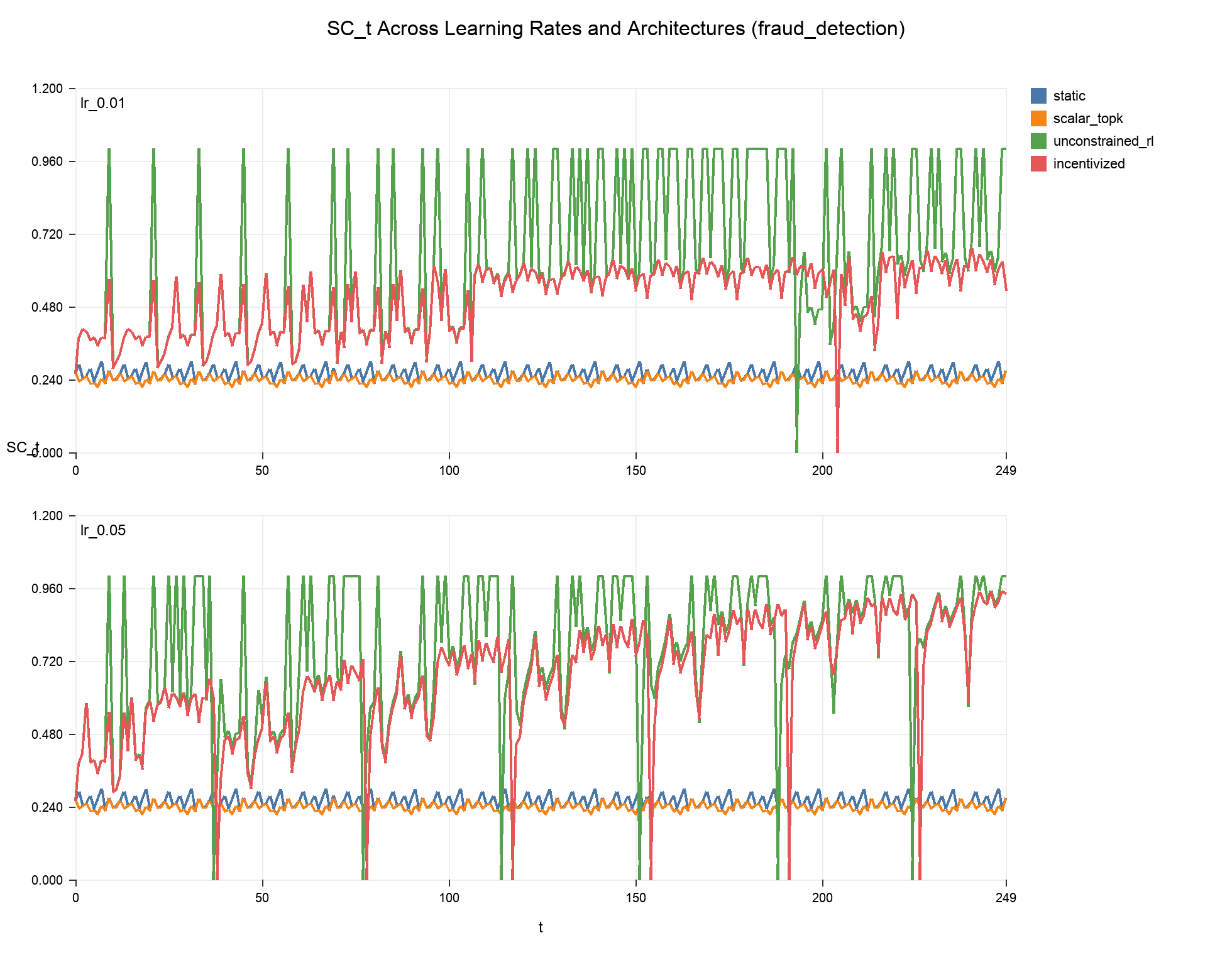}
        \caption{Fraud Detection}
        \label{fig:fraud_sc}
    \end{subfigure}
    \hfill
    \begin{subfigure}[t]{0.32\textwidth}
        \centering
        \includegraphics[width=\linewidth]{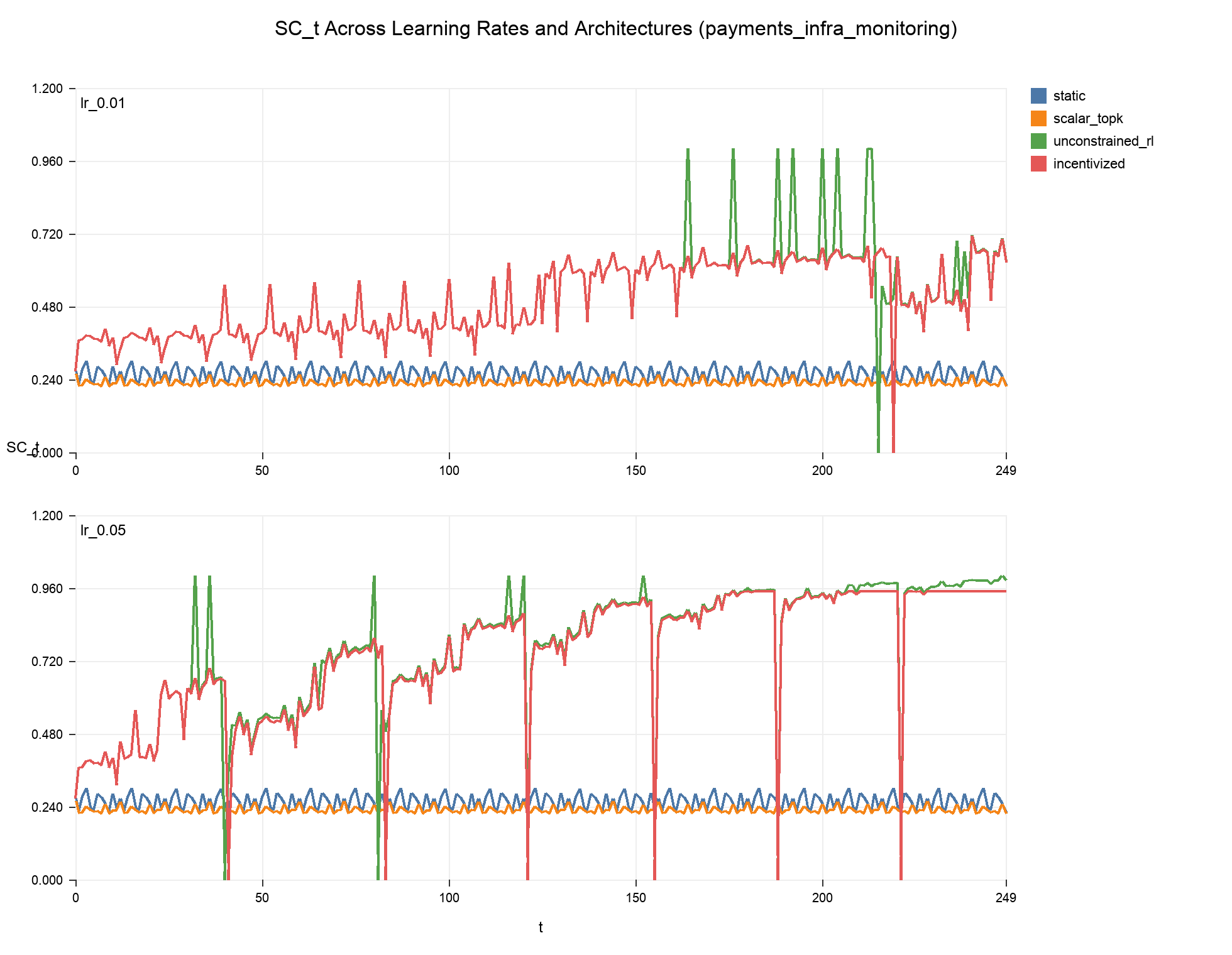}
        \caption{Payments Infrastructure Monitoring}
        \label{fig:payments_sc}
    \end{subfigure}
    \hfill
    \begin{subfigure}[t]{0.32\textwidth}
        \centering
        \includegraphics[width=\linewidth]{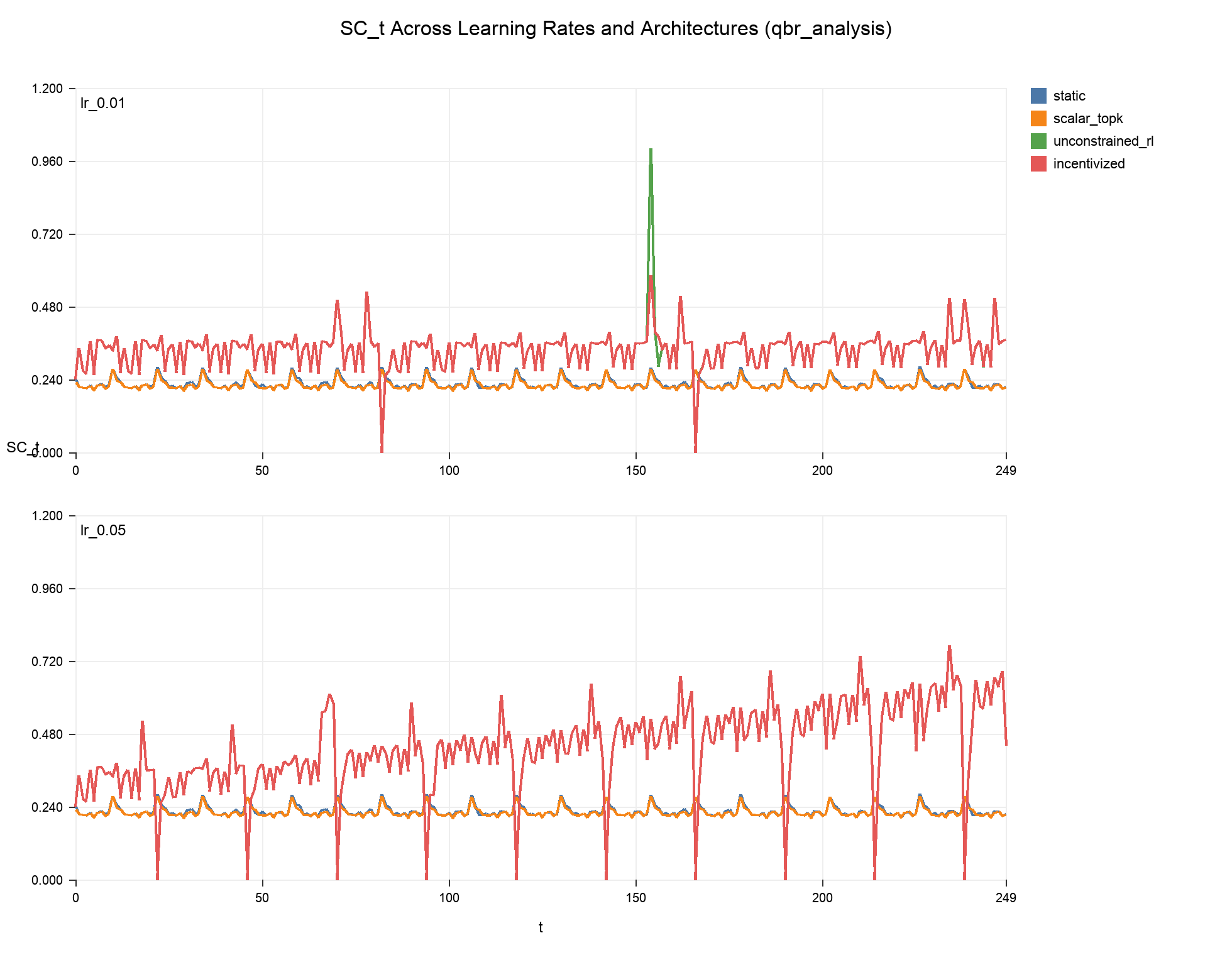}
        \caption{QBR Analysis}
        \label{fig:qbr_sc}
    \end{subfigure}
    
    \caption{
    Temporal evolution of selection concentration 
    ($\mathrm{SC}_t = \max_i P_t(A_i)$) across governance architectures and learning rates.
    Unconstrained reinforcement exhibits rapid concentration growth and, in some cases,
    deterministic lock-in. Incentivized governance allows adaptive reinforcement while
    preserving bounded selection authority through projection-based sovereignty constraints.
    Higher learning rates accelerate reinforcement dynamics but remain constrained under governance.
    }
    
    \label{fig:sc_trajectories}
\end{figure*}

Figures~\ref{fig:fraud_sc}–\ref{fig:qbr_sc} show the temporal evolution of selection concentration $\text{SC}_t$ across learning rates and architectures for each scenario.

Across all domains, unconstrained\_rl exhibits rapid concentration growth, particularly under $lr=0.05$. In fraud detection (Figure~\ref{fig:fraud_sc}), unconstrained\_rl reaches $\text{SC}_T=1.0$, indicating deterministic lock-in under repeated reinforcement.

In contrast, incentivized governance exhibits controlled reinforcement. Concentration increases over time but remains bounded below unity across scenarios. This bounded behavior is visible in Figures~\ref{fig:fraud_sc}, \ref{fig:payments_sc}, and \ref{fig:qbr_sc}, where incentivized trajectories consistently remain below unconstrained reinforcement.

\subsection{Learning Rate Effects}

\begin{figure}[t]
    \centering
    \includegraphics[width=\columnwidth]{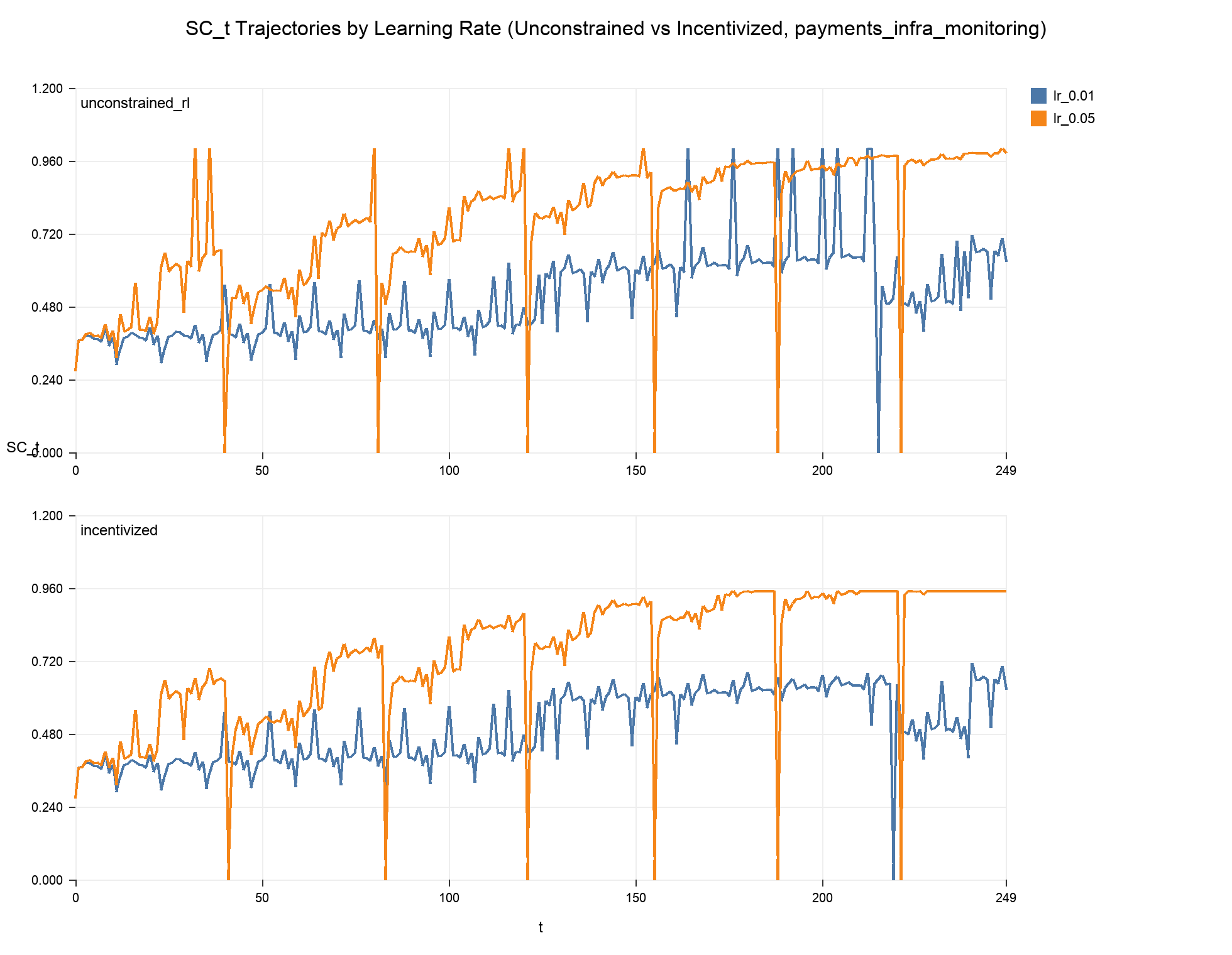}
    \caption{
    Comparison of learning rates ($0.01$ vs $0.05$) on the temporal evolution of selection concentration 
    ($\mathrm{SC}_t$). Higher learning rates accelerate reinforcement dynamics, producing faster growth 
    in concentration. Under unconstrained reinforcement this leads to rapid lock-in, whereas incentivized 
    governance bounds concentration growth through projection-based sovereignty constraints.
    }
    \label{fig:lr_comparison}
\end{figure}

Figure~\ref{fig:lr_comparison} compares $\text{SC}_t$ trajectories under $lr=0.01$ and $lr=0.05$.

Increasing the learning rate accelerates reinforcement dynamics:

\begin{itemize}
    \item In unconstrained\_rl, higher learning rates lead to faster lock-in and greater final concentration.
    \item In incentivized mode, higher learning rates increase concentration but projection prevents deterministic dominance.
\end{itemize}

This demonstrates that learning rate controls reinforcement velocity, while sovereignty projection bounds its magnitude.

\subsection{Realized Selection Dominance (Top-Agent Share)}

\begin{figure*}[t]
    \centering
    \includegraphics[width=0.85\textwidth]{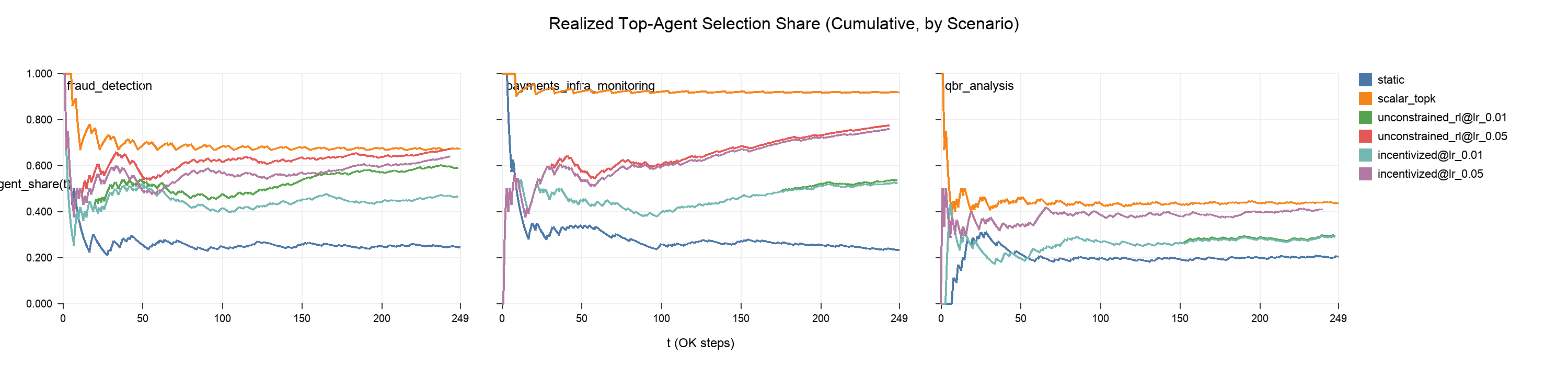}
    \caption{Realized cumulative top-agent selection share across scenarios and governance architectures. Each panel reports the fraction of selections assigned to the most frequently chosen agent as a function of successful steps. Unconstrained reinforcement learning exhibits monotonic convergence toward deterministic dominance, with higher learning rates accelerating lock-in. Scalar top-$k$ produces structurally high dominance from early iterations due to deterministic aggregation. In contrast, incentivized governance increases selection preference for high-performing agents while stabilizing below deterministic collapse through projection-based sovereignty constraints. Static mode remains approximately constant, reflecting the absence of learning.
}
    \label{fig:top_share}
\end{figure*}

While $\text{SC}_t$ measures concentration of the selection distribution, the realized dominance of a single agent is better captured by the empirical selection frequency of the most selected agent, denoted \texttt{top\_agent\_share}. Figure~\ref{fig:top_share} plots this metric over time for each architecture and learning rate.

The patterns are structurally distinct:

\begin{itemize}
    \item \textbf{Unconstrained RL.} Realized dominance increases monotonically and approaches $1.0$ under both learning rates, with $lr=0.05$ converging more rapidly. This confirms deterministic lock-in: once an agent accumulates bias, reinforcement amplifies it without bound.
    
    \item \textbf{Scalar\_topk.} Dominance is high and stable from early iterations. Because the reducer is deterministic and diversity is absent, selection frequency reflects structural monoculture rather than learned adaptation.
    
    \item \textbf{Incentivized Governance.} Dominance increases gradually under reinforcement but stabilizes strictly below unity. Under $lr=0.01$, top-agent share rises slowly and plateaus at moderate levels. Under $lr=0.05$, growth is faster but remains bounded, never reaching deterministic collapse.
    
    \item \textbf{Static.} Selection frequency remains flat and low, reflecting absence of learning.
\end{itemize}

These trajectories reveal a key distinction: projection constraints do not prevent reinforcement-driven preference formation, but they prevent irreversible dominance. Incentivized governance allows adaptive bias accumulation while maintaining residual competition among agents.

\subsection{Governance Debt Trade-off}

Under higher learning rates, incentivized governance incurs measurable governance debt (GD\_dynamic), reflecting projection clipping events when reinforcement attempts to exceed concentration constraints.

This trade-off is visible in Figure~\ref{fig:payments_sc}, where incentivized trajectories approach concentration bounds but are periodically stabilized by projection.

\subsection{Task-Level Reinforcement Dynamics}

The top\_share plots further show that reinforcement effects are most visible under repeated task contexts. In unconstrained\_rl, repeated reward signals quickly drive near-exclusive selection of a single agent. In incentivized governance, repeated contexts still reinforce high-performing agents, but concentration stabilizes at sub-deterministic levels.

Importantly, the gap between unconstrained\_rl and incentivized modes widens with higher learning rates. Under $lr=0.05$, unconstrained reinforcement collapses rapidly, whereas incentivized reinforcement accelerates but remains bounded due to projection clipping.

This empirical divergence confirms that sovereignty projection regulates not only distribution-level concentration (as measured by $\text{SC}_t$), but also realized selection dominance over time.

\subsection{Scenario-Level Metric Summary}
\label{sec:results_table}

Table~\ref{tab:scenario_summary} summarizes the main evaluation metrics across scenarios, governance architectures, and learning rates. Metrics include mean reward, average selection concentration (mean\_SC), initial and final concentration ($\mathrm{SC}_0$, $\mathrm{SC}_T$), reinforced selection concentration (RSC), temporal variance of concentration (var\_SC), and dynamic governance debt (GD\_dynamic).

\begin{table*}[t]
\centering
\caption{Aggregated metrics by scenario, architecture, and learning rate. Values correspond to averages over $T=250$ iterations.}
\label{tab:scenario_summary}
\begin{tabular}{llcccccc}
\hline
Scenario & Mode & LR & mean\_reward & mean\_SC & SC$_T$ & RSC & GD\_dynamic \\
\hline

\multirow{4}{*}{Fraud Detection}
& static & 0.01 & 0.222 & 0.258 & 0.269 & 0.010 & 0.000 \\
& scalar\_topk & 0.01 & 0.422 & 0.239 & 0.268 & 0.009 & 0.000 \\
& unconstrained\_rl & 0.05 & 0.362 & 0.742 & 1.000 & 0.741 & 0.000 \\
& incentivized & 0.05 & 0.358 & 0.669 & 0.942 & 0.683 & 0.004 \\

\hline

\multirow{4}{*}{Payments Monitoring}
& static & 0.01 & 0.165 & 0.256 & 0.222 & -0.047 & 0.000 \\
& scalar\_topk & 0.01 & 0.436 & 0.231 & 0.219 & -0.040 & 0.000 \\
& unconstrained\_rl & 0.05 & 0.374 & 0.755 & 0.985 & 0.716 & 0.000 \\
& incentivized & 0.05 & 0.372 & 0.744 & 0.950 & 0.681 & 0.188 \\

\hline

\multirow{4}{*}{QBR Analysis}
& static & 0.01 & 0.239 & 0.226 & 0.217 & -0.024 & 0.000 \\
& scalar\_topk & 0.01 & 0.288 & 0.223 & 0.215 & -0.016 & 0.000 \\
& unconstrained\_rl & 0.05 & 0.283 & 0.433 & 0.442 & 0.201 & 0.000 \\
& incentivized & 0.05 & 0.283 & 0.433 & 0.442 & 0.201 & 0.000 \\

\hline
\end{tabular}
\end{table*}

\subsection{Summary of Empirical Patterns}
\label{sec:results_summary}

The temporal trajectories shown in Figures~\ref{fig:fraud_sc}–\ref{fig:top_share}, together with the quantitative values reported in Table~\ref{tab:scenario_summary}, reveal a consistent structural pattern across scenarios and learning rates.

First, unconstrained reinforcement produces deterministic concentration. The figures show monotonic growth in both $\mathrm{SC}_t$ and realized top-agent share, while the table confirms that $\mathrm{SC}_T$ approaches or reaches $1.0$ under higher learning rates. Reinforcement without projection therefore induces full lock-in.

Second, incentivized governance enables reinforcement while preserving bounded authority. Across all scenarios, concentration increases under learning but remains strictly below unity. Reinforced Selection Concentration (RSC) remains positive, confirming adaptive improvement, yet final concentration values are consistently lower than in unconstrained\_rl. The visual plateauing observed in the trajectories is numerically reflected in bounded $\mathrm{SC}_T$ values.

Third, learning rate acts as a scaling factor. Increasing the learning rate amplifies reinforcement velocity in both unconstrained and incentivized modes. However, only unconstrained reinforcement collapses into monoculture. In incentivized governance, projection absorbs the increased gradient pressure, converting higher learning rates into faster (but still bounded) adaptation.

Fourth, governance debt operationalizes constraint enforcement. When reinforcement pushes toward prohibited concentration regions, projection clipping occurs, producing measurable governance debt. This effect is visible in scenarios with higher learning rates and confirms that constraint enforcement is active rather than passive.

Finally, deterministic aggregation (scalar\_topk) demonstrates that dominance can arise structurally even without learning. The contrast between scalar\_topk and incentivized modes highlights that governance must regulate both structural aggregation rules and reinforcement dynamics.

Taken together, the figures establish the qualitative dynamics, while the table quantifies their magnitude. The empirical evidence consistently supports the central claim of this work: reinforcement can drive adaptive selection improvement, but only projection-based sovereignty constraints prevent irreversible concentration of decision authority.

%% file: Discussion.tex
\section{Discussion}
\label{sec:discussion}

The experimental results provide evidence that adaptive reinforcement and bounded selection authority can coexist within a single architectural framework. This section interprets the empirical findings in light of the formal model introduced in Section~\ref{sec:problem-setting}.

\subsection{Reinforcement Without Collapse}

A central risk of introducing incentives into selection governance is that reinforcement may reintroduce the very concentration of power that static constraints were designed to prevent. The unconstrained\_rl results demonstrate this failure mode clearly: under iterative reward feedback, selection probability converges toward deterministic dominance, particularly at higher learning rates.

In contrast, incentivized governance exhibits sustained reinforcement while remaining bounded below full concentration. This behavior reflects the effect of projection onto sovereignty constraint sets. Reinforcement updates increase preference for high-performing agents, but projection prevents irreversible collapse.

This empirical divergence supports the core thesis of Selection as Power 2.0:

\begin{quote}
Reinforcement can increase performance without eliminating structural diversity, provided that selection updates are constrained by externally enforced sovereignty projections.
\end{quote}

\subsection{Learning Rate as a Control Parameter}

The learning-rate sweep reveals a structural property of the system: learning rate controls reinforcement velocity, not merely reward accumulation.

Higher learning rates accelerate dominance formation in both unconstrained and incentivized modes. However, the qualitative difference between the two remains: unconstrained reinforcement converges toward monoculture, whereas incentivized reinforcement stabilizes at bounded dominance levels.

This suggests that sovereignty projection transforms learning-rate effects from catastrophic collapse into controlled adaptation.

\subsection{Structural vs Learned Dominance}

The scalar\_topk baseline demonstrates that deterministic dominance does not require learning. Removing diversity and stochastic selection is sufficient to produce structural monoculture.

This distinction is important:

\begin{itemize}
    \item Unconstrained\_rl produces \emph{learned} dominance.
    \item Scalar\_topk produces \emph{structural} dominance.
    \item Incentivized governance mitigates both by preserving diversity constraints.
\end{itemize}

Thus, governance must regulate not only reinforcement dynamics but also structural aggregation rules.

\subsection{Governance Debt and Projection Boundaries}

Projection-based updates introduce governance debt when reinforcement pushes against concentration limits. Empirically, higher learning rates increase governance debt in incentivized mode.

This reflects an inherent trade-off:

\begin{itemize}
    \item Stronger incentives accelerate performance gains.
    \item Stronger incentives increase pressure on sovereignty constraints.
\end{itemize}

Governance debt therefore serves as an operational signal of tension between performance optimization and structural fairness.

Importantly, projection prevents silent drift. Instead of gradual collapse, the system visibly clips updates when constraints are violated. This aligns with the fail-loud principle introduced in the original Selection as Power framework.

\subsection{Dynamic Governance as a Constrained Dynamical System}

From a systems perspective, incentivized selection governance behaves as a constrained dynamical system:

\begin{itemize}
    \item Reward signals introduce directional drift.
    \item Projection introduces reflective boundaries.
    \item Diversity constraints maintain structural heterogeneity.
\end{itemize}

The empirical trajectories illustrate this interaction: reinforcement pushes selection probabilities upward, projection reflects them at constraint boundaries, and diversity prevents collapse into degenerate distributions.

This interpretation connects the empirical results directly to the formal model in Section~\ref{sec:problem-setting}, where updates are projected onto feasible sets $\mathcal{C}$ and $\mathcal{G}$.

\subsection{Limitations of the Present Experiments}

The experiments use a perfect evaluator and synthetic reward structure. While sufficient to demonstrate reinforcement dynamics and constraint effects, they do not capture evaluator noise, delayed rewards, or adversarial feedback.

Additionally, selection concentration is measured at the distribution level ($\mathrm{SC}_t$) and realized frequency level (top-agent share), but long-horizon fairness or regret metrics are not evaluated.

These limitations define the scope of the present claims: the results establish feasibility and boundedness under controlled reinforcement, not universal robustness under all real-world conditions.

\subsection{Summary}

The dynamic experiments support three primary conclusions:

\begin{enumerate}
    \item Reinforcement without constraint produces deterministic dominance.
    \item Projection-based sovereignty constraints bound concentration under iterative updates.
    \item Incentivized governance enables adaptive improvement while preserving structural diversity.
\end{enumerate}

Together, these findings demonstrate that selection governance need not be static. Incentives can be integrated into the architecture without surrendering bounded authority, provided that reinforcement operates strictly within externally enforced constraint sets.

%% file: FutureLines.tex
\section{Limitations and Future Work}
\label{sec:limitations}

While the proposed incentivized governance framework demonstrates bounded adaptive reinforcement under controlled conditions, several limitations constrain the scope and generalizability of the present results.

\subsection{Perfect Evaluator Assumption}

The experiments assume a perfect evaluator that assigns deterministic correctness signals. In practice, evaluators may exhibit noise, bias, or delayed feedback. Under imperfect evaluation, reinforcement updates may amplify misaligned signals, potentially increasing concentration around systematically mis-evaluated agents.

Future work should incorporate stochastic reward models and analyze stability under evaluator noise. In particular, it will be necessary to study whether projection constraints remain sufficient when reward signals are imperfect or adversarial.

\subsection{Synthetic Reward Structure}

The reward function used in the experiments is feature-based and synthetic. While this is sufficient to expose reinforcement and projection dynamics, it does not capture the complexity of real-world reward landscapes. 

In practical deployments, reward signals may depend on long-horizon performance, external regulatory audits, or latent quality metrics. Extending incentivized governance to delayed and partially observable reward settings remains an open problem.

\subsection{Scalability of Constraint Projection}

Projection onto sovereignty constraint sets $\mathcal{C}$ and $\mathcal{G}$ is computationally lightweight in the present experimental configuration. However, as the number of candidate agents or constraint dimensions increases, projection may become more complex.

Future work should explore:

\begin{itemize}
    \item Efficient projection algorithms for high-dimensional parameter spaces,
    \item Adaptive constraint tightening mechanisms,
    \item Theoretical guarantees on projection stability under large-scale systems.
\end{itemize}

\subsection{Learning Rate Sensitivity}

The experiments demonstrate that learning rate strongly influences reinforcement velocity and governance debt. However, a systematic hyperparameter sensitivity analysis was not conducted.

Open questions include:

\begin{itemize}
    \item Whether adaptive learning rate schedules improve stability,
    \item Whether projection thresholds should scale with learning rate,
    \item How governance debt behaves under long-horizon training.
\end{itemize}

Understanding these interactions is critical for practical deployment.

\subsection{Fairness and Long-Horizon Diversity}

The present evaluation measures selection concentration and realized dominance, but does not quantify fairness over extended horizons. Even bounded concentration may produce systematic under-exposure of certain agents.

Future work should incorporate:

\begin{itemize}
    \item Regret-based diversity metrics,
    \item Long-horizon exposure fairness constraints,
    \item Counterfactual fairness evaluation.
\end{itemize}

\subsection{Theoretical Guarantees}

The mathematical analysis in this paper establishes boundedness under projection, but does not provide full convergence guarantees or equilibrium characterization under multi-agent dynamics.

Formal questions remain open:

\begin{itemize}
    \item Under what conditions does constrained reinforcement converge?
    \item Does projection induce oscillatory or chaotic behavior in certain regimes?
\end{itemize}

Bridging empirical stability with formal guarantees represents an important direction for future research.

\subsection{Deployment Considerations}

In real-world regulated systems, governance mechanisms must integrate with audit pipelines, logging requirements, and institutional oversight. The present framework abstracts away deployment constraints such as latency, compute budgets, and cross-system coordination.

Extending incentivized selection governance to production-grade environments will require:

\begin{itemize}
    \item Robust monitoring infrastructure,
    \item Transparent governance reporting,
    \item Institutional calibration of constraint thresholds.
\end{itemize}

\subsection{Summary}

The proposed architecture demonstrates that reinforcement-based adaptation can be bounded through externally enforced sovereignty projections. However, the current evaluation operates under controlled assumptions and simplified reward structures.

Future work should extend the framework to noisy evaluators, adversarial collusion, long-horizon fairness constraints, and formal convergence analysis. Establishing the limits of incentivized governance under realistic deployment conditions remains an open and promising research direction.

%% file: Conclussions.tex
\section{Conclusion}
\label{sec:conclusion}

This work extends the Selection as Power framework from static governance to dynamic, incentivized selection under reinforcement. The central question we addressed was whether adaptive reward-driven updates can coexist with bounded selection authority in high-stakes agentic systems.

The empirical results demonstrate a clear structural distinction. Unconstrained reinforcement leads to deterministic concentration: repeated reward signals amplify selection probability until a single agent dominates. Deterministic aggregation rules (scalar\_topk) produce structural monoculture even without learning. In contrast, incentivized governance, implemented through projection onto externally enforced sovereignty constraint sets, enables adaptive reinforcement while preventing irreversible collapse of diversity.

Learning rate controls the velocity of reinforcement, but projection controls its magnitude. Higher learning rates accelerate preference formation in all learning-enabled architectures; however, only in the absence of projection does this produce deterministic lock-in. When sovereignty constraints are enforced at every update step, concentration growth remains bounded and governance debt reflects active constraint enforcement rather than silent drift.

These findings support a broader claim: selection governance need not be static. Incentive mechanisms can be integrated into the architecture without surrendering bounded authority, provided that reinforcement operates strictly within externally defined constraint sets.

More generally, this work reframes the relationship between incentives and power in autonomous systems. Reinforcement learning is often viewed as a purely performance-oriented mechanism. Our results show that, when embedded within a sovereignty-preserving architecture, reinforcement can improve performance without reintroducing structural concentration of decision authority.

Future work must extend this analysis to noisy evaluators and adversarial reward signals. Formal convergence guarantees under constrained reinforcement remain open. Nonetheless, the present results establish the feasibility of adaptive selection governance: bounded authority can be preserved even under sustained reinforcement pressure.

In regulated and high-stakes environments, governance must regulate not only what agents can select, but how reinforcement reshapes that authority over time. Selection as Power 2.0 provides a principled step toward that goal.